\documentclass[conference]{IEEEtran}
\usepackage{graphicx}
\usepackage{amsmath}
\usepackage{amssymb}
\usepackage{mathrsfs}
\usepackage{stfloats}
\usepackage{dsfont}
\usepackage{setspace}
\usepackage{epstopdf}
\usepackage{cite}
\usepackage{balance}
\newtheorem{proposition}{Proposition}
\newtheorem{lemma}{Lemma}

\begin{document}
\title{MIMO Cellular Networks with Simultaneous Wireless Information and Power Transfer}
\author{
\authorblockN{Tu Lam Thanh$^{(1)}$, Marco Di Renzo$^{(1)}$, and Justin P. Coon$^{(2)}$}
\authorblockA{$^{(1)}$ Laboratoire des Signaux et Syst\`emes, CNRS, CentraleSup\'elec, Univ Paris Sud\\
Universit\'e Paris-Saclay \\
3 rue Joliot Curie, Plateau du Moulon, 91192, Gif-sur-Yvette, France \\
e-mail: \{lamthanh.tu, marco.direnzo\}@l2s.centralesupelec.fr}
\authorblockA{$^{(2)}$ Oxford University -- Department of Engineering Science \\
Parks Road, Oxford, OX1 3PJ, United Kingdom \\
e-mail: justin.coon@eng.ox.ac.uk} \vspace{-0.5cm}
}
%
\maketitle
\begin{abstract}
In this paper, we introduce a mathematical approach for system-level analysis and optimization of densely deployed multiple-antenna cellular networks, where low-energy devices are capable of decoding information data and harvesting power simultaneously. The base stations are assumed to be deployed according to a Poisson point process and tools from stochastic geometry are exploited to quantify the trade-off in terms of information rate and harvested power. It is shown that multiple-antenna transmission is capable of increasing information rate and harvested power at the same time.
\end{abstract}
\section{Introduction} \label{Introduction}
The Internet of Things (IoT) is expected to connect billions of Low-energy Devices (LeDs) by 2020 \cite{IoT_Survey2015}. One of the main challenges of the IoT is how to provide enough energy for the electronics of the LeDs, in order to have them operational over a reasonable amount of time without making their battery too large or the device itself too bulky. For several applications, it may not be even possible to (re-)charge some LeDs.

In this context, the emerging concept of Simultaneous Wireless Information and Power Transfer (SWIPT) may constitute a suitable solution for prolonging the battery life of LeDs and, in a foreseeable future, for making them energy-neutral, \textit{i.e.}, operational in a complete self-powered fashion. SWIPT, in particular, is a technology where the same radio frequency signal is used for information transmission and for replenishing the batteries of the LeDs \cite{Krikidis__COMMAG2014}. SWIPT may find application in the emerging market of cellular-enabled IoT, where LeDs, \textit{e.g.}, smart-watches \cite{INTEL_MICA}, receive notifications from their own cellular connection \cite{ATT} and, simultaneously, re-charge their battery. The recent decision to standardize NarrowBand IoT (NB-IoT), a new narrow-band radio technology that addresses the requirements of the IoT, confirms the wish of capitalizing on the ubiquitous coverage offered by the cellular network infrastructure for IoT applications \cite{NB_IOT}.

In spite of these potential advantages, the design and optimization of SWIPT-enabled cellular networks, which provide data connectivity and power to LeDs, pose new challenges and introduce never observed performance trade-offs. Cellular networks are designed based on the assumption that the other-cell interference has a negative impact on Wireless Information Transfer (WIT), since it reduces the coverage probability and average rate \cite{MDR_TCOMrate}. Interference, on the other hand, constitutes a natural source to be exploited for enabling Wireless Power Transfer (WPT) \cite{Zheng__COMMAG2014}. In SWIPT-enabled cellular networks, as a result, the development of interference management techniques that exploit interference for WPT and counteract it for WIT plays a fundamental role, especially if the LeDs operate in scenarios with adverse networking conditions.

The application of smart antenna technologies, \textit{i.e.}, Multiple-Input-Multiple-Output (MIMO) transmission and reception schemes, constitutes a promising solution for managing and exploiting the interference in SWIPT-enabled cellular networks \cite{Schober__COMMAG2015}. More specifically, MIMO schemes can be exploited to provide two distinct benefits. On the one hand, multiple antennas at the receiver can be used for increasing the amount of harvested power via spatial diversity. On the other hand, multiple antennas at the transmitter can be used for improving the efficiency of information and energy transfer via spatial beamforming. A summary of recent results on the application of smart antenna technologies to SWIPT-enabled systems is available in \cite{SWIPTmimo__COMMAG2015}. Most of the existing works, however, are applicable to small-scale network topologies. Large-scale networks are, on the other hand, much less investigated. In \cite{Krikidis__Mar2014} and \cite{Ding__Aug2014}, relay-aided networks are studied. In \cite{Durrani__WCLAug2015} and \cite{Flint__TWC2015} ad hoc networks are analyzed. These papers, however, consider single-antenna transmission. In \cite{Durrani__WCL2015}, ad hoc networks with multiple-antenna transmitters are studied. The analysis, however, is not directly applicable to cellular networks, since the spatial constraints imposed by the cell association are not taken into account. The only paper where the performance of SWIPT-enabled cellular networks is investigated is \cite{MDR_Globecom2015}, which, however, does not consider MIMO either at the Base Stations (BSs) or at the LeDs.

In this paper, motivated by these considerations, we introduce a tractable mathematical approach for quantifying the benefits, in terms of information rate and harvested power, of using MIMO in SWIPT-enabled cellular networks. In particular, Maximum Ratio Transmission (MRT) and Maximum Ratio Combining (MRC) at the BSs and at the LeDs are considered, respectively. The proposed approach relies on modeling the locations of the BSs are points of a Poisson Point Process (PPP) and uses stochastic geometry for system-level analysis. It accounts for the spatial correlation of the other-cell interference that emerges from SWIPT operation and is applicable to BSs and LeDs equipped with an arbitrary number of transmit and receive antennas, respectively. With the use of MIMO, it is proved that information rate and harvested power can be improved simultaneously.

The paper is organized as follows. In Section \ref{SystemModel}, the system model is introduced. In Section \ref{MIMO}, MRT and MRC schemes are described. In Section \ref{JCCDF}, the trade-off between information rate and harvested power is quantified in terms of their Joint Complementary Cumulative Distribution Function (J-CCDF). In Section \ref{Results}, analysis and findings are validated with the aid of numerical simulations. Section \ref{Conclusion} concludes this paper.

\emph{Notation}:
%
%
$X \sim \mathcal{CN} \left( \mu, \sigma^2 \right)$ denotes that $X$ is a complex Gaussian random variable with zero mean and variance $\sigma^2$.
$X \sim {\mathcal{E}}\left( \Omega \right)$ denotes that $X$ is an exponential random variable with mean $\Omega$.
%
%
$j = \sqrt { - 1}$ is the imaginary unit.
${\mathbb{E}}\left\{  \cdot  \right\}$ is the expectation operator.
${\left( \cdot \right)!}$ is the factorial operator.
%
%
${\mathop{\rm Im}\nolimits} \left\{  \cdot  \right\}$ denotes the imaginary part.
$g^{\left( n \right)} \left(  \cdot  \right)$ denotes the $n$th derivative of $g \left(  \cdot  \right)$ with respect to its argument.
${\bf{1}}\left(  \cdot  \right)$ is the indicator function.
${\mathcal{H}}\left(  \cdot  \right)$ is the Heaviside function and ${\mathcal{\overline{H}}}\left( x \right) = 1 - {\mathcal{H}}\left( x \right)$.
${}_p{F_q}\left( {{a_1},\ldots,{a_p};{b_1},\ldots,{b_q}; \cdot} \right)$ is the generalized hypergeometric function \cite[Eq. 9.14.1]{GradshteynRyzhik}.
$\Gamma \left( {\cdot,\cdot} \right)$ is the upper-incomplete Gamma function \cite[Eq. 8.350.2]{GradshteynRyzhik}.
$\min \left\{ { \cdot , \cdot } \right\}$ and $\max \left\{ { \cdot , \cdot } \right\}$ are the minimum and the maximum functions, respectively.
$f_X \left(  \cdot  \right)$, $F_X \left(  \cdot  \right)$ and $\Phi_X \left(  \cdot  \right)$ are the Probability Density Function (PDF), the Cumulative Distribution Function (CDF) and the Characteristic Function (CF) of the random variable $X$.
\section{System Model} \label{SystemModel}
\subsection{Cellular Networks Modeling}
A downlink MIMO cellular network is considered, where the BSs and the LeDs are equipped with $N_t$ and $N_r$ antennas, respectively. The BSs are modeled as points of a homogeneous PPP, denoted by $\Psi$, of density $\lambda$. They have fixed transmit power equal to $P$. Without loss of generality, the analysis is performed for the typical LeD that is located at the origin \cite{MDR_TCOMrate}. This typical LeD is served by the BS that provides the smallest path-loss, while all the other BSs act as interferers. Throughout this paper, the superscripts, subscripts and indexes ``0'', ``i'' and ``n'' are referred to the serving BS, to the generic interfering BS, and to the generic BS, respectively.
\subsection{SWIPT Based on Power Splitting (PS)}
The typical LeD is equipped with an information and an energy receivers that operate according to the PS scheme \cite{Krikidis__COMMAG2014}. Let ${\rm{P}}_{{\rm{RX}}}$ be the received power at the LeD. It is split in two parts: ${\rm{P}}_{{\rm{EH}}}  = \rho {\rm{P}}_{{\rm{RX}}}$ is used for Energy Harvesting (EH) and ${\rm{P}}_{{\rm{ID}}}  = {\rm{P}}_{{\rm{RX}}}  - {\rm{P}}_{{\rm{EH}}}  = \left( {1 - \rho } \right){\rm{P}}_{{\rm{RX}}}$ is used for Information Decoding (ID), where $0 \le \rho  \le 1$ is the power splitting ratio. The proposed approach can be generalized for application to the Time Switching (TS) scheme \cite{Krikidis__COMMAG2014}, as discussed in \cite{MDR_Globecom2015}.
\subsection{Channel Modeling}
A realistic channel model is considered, which includes Line-of-Sight (LOS) and Non-LOS (NLOS) links due to the presence of spatial blockages, the distance-dependent path-loss, and the fast-fading. Shadowing is implicitly taken into account by the LOS/NLOS link model. The adopted model is in agreement with \cite{ACM_MDR}, which has been experimentally validated for application to dense urban environments.

\subsubsection{LOS/NLOS Links} Let $r$ be the distance from a generic BS to the typical LeD. The probabilities of occurrence $p_{{\rm{LOS}}} \left(  \cdot  \right)$ and $p_{{\rm{NLOS}}} \left(  \cdot  \right)$ of LOS and NLOS links, respectively, as a function of $r$, are formulated as follows:
\begin{equation}
\label{Eq_1}
p_{s} \left( r \right) = \begin{cases}
 q_{s}^{\left[ {0,D} \right]} & \quad {\rm{if}}\quad r \in \left[ {0,D} \right) \\
 q_{s}^{\left[ {D,\infty } \right]} & \quad {\rm{if}}\quad r \in \left[ {D, + \infty } \right) \\
 \end{cases}
\end{equation}
\noindent where $q_{s}^{\left[ {a,b} \right]}$ for $s \in \left\{ {{\rm{LOS}},{\rm{NLOS}}} \right\}$ denotes the probability that a link of length $r \in \left[ {a,b} \right)$ is in state $s$. Since a link can only be in LOS or in NLOS, the equality $q_{{\rm{LOS}}}^{\left[ {a,b} \right]} + q_{{\rm{NLOS}}}^{\left[ {a,b} \right]}  = 1$ holds. The parameter $D$ is a breaking distance that takes into account that the probability of LOS and NLOS is usually different for short and long transmission distances.

By assuming no spatial correlation between the links, $\Psi$ can be partitioned in two independent and non-homogeneous PPPs, denoted by $\Psi _{{\rm{LOS}}}$ and $\Psi _{{\rm{NLOS}}}$, such that $\Psi  = \Psi _{{\rm{LOS}}}  \cup \Psi _{{\rm{NLOS}}}$. From \eqref{Eq_1}, the densities of $\Psi _{{\rm{LOS}}}$ and $\Psi _{{\rm{NLOS}}}$ are $\lambda _{{\rm{LOS}}} \left( r \right) = \lambda p_{{\rm{LOS}}} \left( r \right)$ and $\lambda _{{\rm{NLOS}}} \left( r \right) = \lambda p_{{\rm{NLOS}}} \left( r \right)$, respectively.
\subsubsection{Path-Loss}
The path-loss of LOS and NLOS links is $l_{s} \left( r \right) = \kappa _{s} r^{\beta _{s} }$ for $s \in \left\{ {{\rm{LOS}},{\rm{NLOS}}} \right\}$, where $\kappa _{{s}}  = \left( {{{4\pi } \mathord{\left/ {\vphantom {{4\pi } \nu }} \right. \kern-\nulldelimiterspace} \nu }} \right)^{2}$ is the pathloss constant, $\nu$ is the transmission wavelength, and $\beta _{s}$ is the power path-loss exponent.
\subsubsection{Fast-Fading}
All channels are assumed to be independent and identically distributed (i.i.d.) complex Gaussian random variables with zero mean and unit variance.
\subsection{Cell Association} \label{CellAssocitation}
The typical LeD is served by the BS that provides the smallest path-loss. This smallest path-loss can be formulated as $L^{\left( 0 \right)}  = \min \left\{ {L_{{\rm{LOS}}}^{\left( 0 \right)} ,L_{{\rm{NLOS}}}^{\left( 0 \right)} } \right\}$, where $L_{{s}}^{\left( 0 \right)}$ for $s \in \left\{ {{\rm{LOS}},{\rm{NLOS}}} \right\}$ is the smallest path-loss of LOS/NLOS links:
\begin{equation}
\label{Eq_2}
L_{{\mathop{s}\nolimits} }^{\left( 0 \right)}  = \mathop {\min }\limits_{n \in \Psi _{{s}} } \left\{ {l_{{s}} \left( {r^{\left( n \right)} } \right)} \right\}
\end{equation}
\noindent and $r^{(n)}$ is the distance between a generic BS and the LeD.
\begin{figure*}[!t]
\setcounter{equation}{11}
\begin{equation}
\label{Eq_12}
{\Lambda _s}\left( {\left[ {0,x} \right)} \right) = \pi \lambda q_s^{\left[ {0,D} \right]}{\left( {{x \over {{\kappa _s}}}} \right)^{{2 \over {{\beta _s}}}}}\overline {\cal H}\left( {x - {\kappa _s}{D^{{\beta _s}}}} \right) + \pi \lambda \left( {{{\left( {{x \over {{\kappa _s}}}} \right)}^{{2 \over {{\beta _s}}}}}q_s^{\left[ {D,\infty } \right]} + {D^2}\left( {q_s^{\left[ {0,D} \right]} - q_s^{\left[ {D,\infty } \right]}} \right)} \right){\cal H}\left( {x - {\kappa _s}{D^{{\beta _s}}}} \right)
\end{equation}
\normalsize \hrulefill \vspace*{-0pt}
\end{figure*}
\begin{figure*}[!t]
\setcounter{equation}{12}
\begin{equation}
\label{Eq_13}
\Lambda _s^{\left( 1 \right)} \left( {\left[ {0,x} \right)} \right) = \left( {{{2\pi \lambda } \mathord{\left/
 {\vphantom {{2\pi \lambda } {\beta _s }}} \right.
 \kern-\nulldelimiterspace} {\beta _s }}} \right)\kappa _s^{ - {2 \mathord{\left/
 {\vphantom {2 {\beta _s }}} \right.
 \kern-\nulldelimiterspace} {\beta _s }}} x^{\left( {{2 \mathord{\left/
 {\vphantom {2 {\beta _s }}} \right.
 \kern-\nulldelimiterspace} {\beta _s }} - 1} \right)} \left( {q_s^{\left[ {0,D} \right]} {\mathcal{\overline H}}\left( {x - \kappa _s D^{\beta _s } } \right) + q_s^{\left[ {D,\infty } \right]} {\mathcal{H}}\left( {x - \kappa _s D^{\beta _s } } \right)} \right)
\end{equation}
\normalsize \hrulefill \vspace*{-0pt}
\end{figure*}
\begin{figure*}[!t]
\setcounter{equation}{14}
\begin{equation}
\label{Eq_15}
\begin{split}
 \Phi_{\mathcal{I}}  \left( {\left. \omega  \right|L^{\left( 0 \right)} ;s} \right) & = \exp \left( {\lambda \pi q_s^{\left[ {D,\infty } \right]} \max \left\{ {D^2 ,\left( {{{L^{\left( 0 \right)} }}/{{\kappa _s }}} \right)^{{2 \mathord{\left/
 {\vphantom {2 {\beta _s }}} \right.
 \kern-\nulldelimiterspace} {\beta _s }}} } \right\}\left( {1 - \Upsilon _s \left( {\omega ,\max \left\{ {\kappa _s D^{\beta _s } ,L^{\left( 0 \right)} } \right\}} \right)} \right)} \right) \\
 & \hspace{-0.5cm} \times \exp \left( {\pi \lambda q_s^{\left[ {0,D} \right]} \left[ {\left( {{{L^{\left( 0 \right)} }}/{{\kappa _s }}} \right)^{{2 \mathord{\left/
 {\vphantom {2 {\beta _s }}} \right.
 \kern-\nulldelimiterspace} {\beta _s }}} \left( {1 - \Upsilon _s \left( {\omega ,L^{\left( 0 \right)} } \right)} \right) - D^2 \left( {1 - \Upsilon _s \left( {\omega ,\kappa _s D^{\beta _s } } \right)} \right)} \right]\overline {\mathcal{H}} \left( {L^{\left( 0 \right)}  - \kappa _s D^{\beta _s } } \right)} \right)
 \end{split}
\end{equation}
\normalsize \hrulefill \vspace*{-0pt}
\end{figure*}
\begin{figure*}[!t]
\setcounter{equation}{16}
\begin{equation}
\label{Eq_17}
\begin{split}
 & {\mathcal{J}}_{v,u}^{\left( 1 \right)}  = \int\nolimits_0^{ + \infty } {\int\nolimits_0^{ + \infty } {\frac{1}{{\pi \omega }}{\mathop{\rm Im}\nolimits} \left\{ {\exp \left( { - j\omega \frac{{q_* }}{P}} \right)\left( {v - \frac{{j\omega }}{y}} \right)^{ - \left( {1 + u} \right)} \Gamma \left( {1 + u,\frac{{{\mathcal{T}}_* }}{P}\left( {vy - j\omega } \right)} \right)\Phi _{\mathcal{I}} \left( {\left. \omega  \right|y} \right)} \right\}f_{L^{\left( 0 \right)} } \left( y \right)d\omega dy} }  \\
 & {\mathcal{J}}_{v,u}^{\left( 2 \right)}  = \int\nolimits_0^{ + \infty } {\int\nolimits_0^{ + \infty } {\frac{1}{{\pi \omega }}{\mathop{\rm Im}\nolimits} \left\{ {\exp \left( {j\omega \frac{{\sigma _*^2 }}{P}} \right)\left( {v + \frac{{j\omega r_* }}{y}} \right)^{ - \left( {1 + u} \right)} \Gamma \left( {1 + u,\frac{{{\mathcal{T}}_* }}{P}\left( {vy + j\omega r_* } \right)} \right)\Phi _{\mathcal{I}} \left( {\left. \omega  \right|y} \right)} \right\}f_{L^{\left( 0 \right)} } \left( y \right)d\omega dy} }
 \end{split}
\end{equation}
\normalsize \hrulefill \vspace*{-0pt}
\end{figure*}
\section{Information Rate and Harvested Power Using MRT and MRC} \label{MIMO}
We assume that MRT and MRC are employed at the BSs and at the LeDs \cite{MIMO_Slim}, \cite{MIMO_Sonia}, respectively. Similar to \cite{MDR_Globecom2015}, the Shannon rate (in bits/sec), $\mathcal{R}$, of the ID receiver and the average harvested power (in Watt), $\mathcal{Q}$, of the Energy Harvesting (EH) receiver can be formulated as follows:
\begin{equation}
\label{Eq_7}
\begin{split}
& {\mathcal{R}} = B_w \log _2 \left( {1 + \frac{{{{P\chi ^{\left( 0 \right)} } \mathord{\left/
 {\vphantom {{P\chi ^{\left( 0 \right)} } {L^{\left( 0 \right)} }}} \right.
 \kern-\nulldelimiterspace} {L^{\left( 0 \right)} }}}}{{P{\mathcal{I}}_{{\rm{ID}}}  + \sigma _N^2  + {{\sigma _{{\rm{ID}}}^2 } \mathord{\left/
 {\vphantom {{\sigma _{{\rm{ID}}}^2 } {\left( {1 - \rho } \right)}}} \right.
 \kern-\nulldelimiterspace} {\left( {1 - \rho } \right)}}}}} \right) \\
& {\mathcal{Q}} = \rho \zeta \left( {{{P\chi ^{\left( 0 \right)} } \mathord{\left/
 {\vphantom {{P\chi ^{\left( 0 \right)} } {L^{\left( 0 \right)} }}} \right.
 \kern-\nulldelimiterspace} {L^{\left( 0 \right)} }} + P{\mathcal{I}}_{{\rm{EH}}} } \right)
\end{split}
\end{equation}
\noindent where $B_w$ is the transmission bandwidth, $0 \le \zeta \le 1$ is the conversion efficiency of the EH receiver, and $\sigma _{{\rm{ID}}}^2$ is the variance of the additive white Gaussian noise of the ID receiver, which is due to the conversion of the received signal from radio frequency to baseband.

In addition, ${\chi ^{\left( 0 \right)} }$ is the power gain of the intended link that results from using MRT at the serving BS and MRC at the typical LeD, respectively. From \cite[Eq. (9)]{MIMO_Sonia}, its PDF can be formulated as follows:
\begin{equation}
\label{Eq_5}
f_{\chi ^{\left( 0 \right)} } \left( \xi  \right) = {\mathcal{K}}_{p,q} \sum\limits_{v = 1}^q {\sum\limits_{u = p - q}^{\left( {p + q - 2v} \right)v} {c_{v,u} \xi ^u \exp \left( { - v\xi } \right)} }
\end{equation}
\noindent where $p = \max \left\{ {N_t ,N_r } \right\}$, $q = \min \left\{ {N_t ,N_r } \right\}$, ${\mathcal{K}}_{p,q}  = \left( {\prod\nolimits_{a = 1}^q {\left( {q - a} \right)!\left( {p - a} \right)!} } \right)^{ - 1}$, and the coefficients $c_{v,u}$ are obtained by using \cite[Algorithm 1]{MIMO_Sonia}.

Finally, ${\mathcal{I}} = {\mathcal{I}}_{{\rm{ID}}}  = {\mathcal{I}}_{{\rm{EH}}}$ is the aggregate other-cell interference, which for MRT/MRC can be formulated as \cite{MIMO_Slim}:
\begin{equation}
\label{Eq_8}
\begin{split}
 {\mathcal{I}} & = \sum\limits_{i \in \Psi _{{\rm{LOS}}} } {\left( {{1 \mathord{\left/
 {\vphantom {1 {l_{{\rm{LOS}}} \left( {r^{\left( i \right)} } \right)}}} \right.
 \kern-\nulldelimiterspace} {l_{{\rm{LOS}}} \left( {r^{\left( i \right)} } \right)}}} \right)\gamma ^{\left( i \right)} {\bf{1}}\left( {l_{{\rm{LOS}}} \left( {r^{\left( i \right)} } \right) > L^{\left( 0 \right)} } \right)}  \\
 & + \sum\limits_{i \in \Psi _{{\rm{NLOS}}} } {\left( {{1 \mathord{\left/
 {\vphantom {1 {l_{{\rm{NLOS}}} \left( {r^{\left( i \right)} } \right)}}} \right.
 \kern-\nulldelimiterspace} {l_{{\rm{NLOS}}} \left( {r^{\left( i \right)} } \right)}}} \right)\gamma ^{\left( i \right)} {\bf{1}}\left( {l_{{\rm{NLOS}}} \left( {r^{\left( i \right)} } \right) > L^{\left( 0 \right)} } \right)}
\end{split}
\end{equation}
\noindent where $\gamma ^{\left( i \right)} \sim {\mathcal{E}}\left( 1 \right)$ is the gain of the $i$th interfering link. In particular, $\gamma ^{\left( i \right)}$ and ${\chi ^{\left( 0 \right)} }$ are independent of each other \cite{MIMO_Slim}.
\section{System-Level Analysis of Information Rate and Harvested Power} \label{JCCDF}
In this section, our main result is introduced. We develop, in particular, a mathematical framework that quantifies the trade-off between information rate and harvested power. This trade-off is quantified with the aid of the J-CCDF of $\mathcal{R}$ and $\mathcal{Q}$ in \eqref{Eq_7}. In mathematical terms, the J-CCDF is as follows:
\setcounter{equation}{8}
\begin{equation}
\label{Eq_9}
F_c \left( {{\mathcal{R}}_* ,{\mathcal{Q}}_* } \right) = \Pr \left\{ {{\mathcal{R}} \ge {\mathcal{R}}_* ,{\mathcal{Q}} \ge {\mathcal{Q}}_* } \right\}
\end{equation}
\noindent where ${\mathcal{R}}_{*}$ and ${\mathcal{Q}}_{*}$ represent the minimum information rate and harvested power that need to be guaranteed for the LeD being able to perform its tasks. The mathematical characterization of $F_c \left( \cdot, \cdot \right)$ provides complete information on the achievable performance of SWIPT-enabled cellular networks.

To facilitate the computation of the J-CCDF in \eqref{Eq_9}, we introduce two lemmas that provide the PDF and the CDF of the smallest path-loss, ${L^{\left( 0 \right)} }$, and the CF of the aggregate other-cell interference, ${\mathcal{I}}$, conditioned on ${L^{\left( 0 \right)} }$. Their proofs can be obtained by using similar mathematical steps as those reported in \cite{MDR_Globecom2015}. Due to space limitations, the details are omitted.
\begin{lemma} \label{CDF_PDF__L0}
The CDF and PDF of the smallest path-loss, $L^{\left( 0 \right)}$, defined in Section \ref{CellAssocitation} are as follows:
\begin{equation}
\label{Eq_10}
\begin{split}
& F_{L^{(0)} } \left( x \right) = 1 - \exp \left( { - \Lambda \left( {\left[ {0,x} \right)} \right)} \right) \\
& f_{L^{(0)} } \left( x \right) = \Lambda ^{\left( 1 \right)} \left( {\left[ {0,x} \right)} \right)\exp \left( { - \Lambda \left( {\left[ {0,x} \right)} \right)} \right)
\end{split}
\end{equation}
\noindent where the following definitions hold:
\begin{equation}
\label{Eq_11}
\begin{split}
& \Lambda \left( {\left[ {0,x} \right)} \right) = \Lambda _{{\mathop{\rm LOS}\nolimits} } \left( {\left[ {0,x} \right)} \right) + \Lambda _{{\rm{NLOS}}} \left( {\left[ {0,x} \right)} \right) \\
& \Lambda^{\left( 1 \right)} \left( {\left[ {0,x} \right)} \right) = \Lambda _{{\rm{LOS}}}^{\left( 1 \right)} \left( {\left[ {0,x} \right)} \right) + \Lambda _{{\rm{NLOS}}}^{\left( 1 \right)} \left( {\left[ {0,x} \right)} \right)
\end{split}
\end{equation}
\noindent as well as $\Lambda _s \left( {\left[ {\cdot,\cdot} \right)} \right)$ and $\Lambda _s^{\left( 1 \right)} \left( {\left[ { \cdot , \cdot } \right)} \right)$ for $s \in \left\{ {{\rm{LOS}},{\rm{NLOS}}} \right\}$ are defined in \eqref{Eq_12} and \eqref{Eq_13} shown at the top of this page.

\emph{Proof}: See \cite{MDR_Globecom2015}. \hfill $\Box$
\end{lemma}
\begin{lemma} \label{CF__L0}
Given $L^{\left( 0 \right)}$, the CF of of the aggregate other-cell interference, ${\mathcal{I}}$, is as follows:
\setcounter{equation}{13}
\begin{equation}
\label{Eq_14}
\Phi_{\mathcal{I}} \left( {\left. \omega  \right|L^{\left( 0 \right)} } \right) = \Phi_{\mathcal{I}} \left( {\left. \omega  \right|L^{\left( 0 \right)} ;{\rm{LOS}}} \right)\Phi_{\mathcal{I}} \left( {\left. \omega  \right|L^{\left( 0 \right)} ;{\rm{NLOS}}} \right)
\end{equation}
\noindent where $\Phi_{\mathcal{I}} \left( {\left. \cdot  \right|L^{\left( 0 \right)} ;s} \right)$ for $s \in \left\{ {{\rm{LOS}},{\rm{NLOS}}} \right\}$ is defined in \eqref{Eq_15} shown at the top of the previous page and ${\Upsilon_s}\left( {\omega ,Z} \right) = {\;_2}{F_1}\left( {1, - {2}/{{{\beta_s}}},1 - {2}/{{{\beta_s}}},{{j\omega }}/{{Z}}} \right)$.

\emph{Proof}: See \cite{MDR_Globecom2015}. \hfill $\Box$
\end{lemma}

From \textit{Lemma \ref{CDF_PDF__L0}} and \textit{Lemma \ref{CF__L0}}, an exact and tractable expression of the J-CCDF in \eqref{Eq_9} is given in \textit{Proposition \ref{JCCDF__Prop}}.
\begin{proposition} \label{JCCDF__Prop}
The J-CCDF in \eqref{Eq_9} can be formulated as:
\setcounter{equation}{15}
\begin{equation}
\label{Eq_16}
F_c \left( {{\mathcal{R}}_* ,{\mathcal{Q}}_* } \right) = {\mathcal{K}}_{p,q} \sum\limits_{v = 1}^q {\sum\limits_{u = p - q}^{\left( {p + q - 2v} \right)v} {c_{v,u} \left( {{\mathcal{J}}_{v,u}^{\left( 1 \right)}  - {\mathcal{J}}_{v,u}^{\left( 2 \right)} } \right)} }
\end{equation}
\noindent where ${{\mathcal{J}}_{v,u}^{\left( 1 \right)} }$ and ${{\mathcal{J}}_{v,u}^{\left( 2 \right)} }$ are provided in \eqref{Eq_17} shown at the top of the previous page, as well as $r_*  = \left( {2^{{{{\mathcal{R}}_* } \mathord{\left/ {\vphantom {{{\mathcal{R}}_* } {B_w }}} \right. \kern-\nulldelimiterspace} {B_w }}}  - 1} \right)^{ - 1}$, $\sigma _*^2  = \sigma _N^2  + \sigma _{{\rm{ID}}}^2 \left( {1 - \rho } \right)^{ - 1}$, $q_*  = {\mathcal{Q}}_* \left( {\rho \zeta } \right)^{ - 1}$ and ${{{\mathcal{T}}_*  = \left( {q_*  + \sigma _*^2 } \right)} \mathord{\left/ {\vphantom {{{\mathcal{T}}_*  = \left( {q_*  + \sigma _*^2 } \right)} {\left( {r_*  + 1} \right)}}} \right. \kern-\nulldelimiterspace} {\left( {r_*  + 1} \right)}}$.

\emph{Proof}: See Appendix \ref{Appendix}. \hfill $\Box$
\end{proposition}

The J-CCDF in \eqref{Eq_16} is formulated in terms of a two-fold integral that can be efficiently computed with the aid of state-of-the-art computational software programs and has the advantage of avoiding lengthly Monte Carlo simulations.
\begin{figure}[!t]
\centering
\includegraphics [width=1.0\columnwidth] {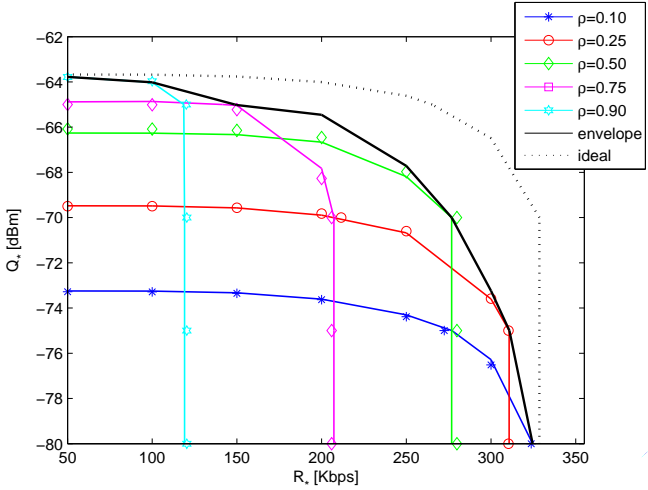}
\vspace{-0.75cm} \caption{Contour lines of the J-CCDF as a function of $\rho$ that correspond to the pairs $\left( {{\mathcal{R}}_* ,{\mathcal{Q}}_* } \right)$ such that $F_c \left( {{\mathcal{R}}_* ,{\mathcal{Q}}_* } \right) = 0.75$. The solid lines show the mathematical framework in \eqref{Eq_16} and the markers show Monte Carlo simulations. The curve ``envelope'' corresponds to the setup of $\rho$ that provides the best J-CCDF. The curve ``ideal'' represents the setup where EH and ID can be performed simultaneously. Setup: $N_t=4$ and $N_r=2$.} \label{Fig_1} \vspace{-0.25cm}
\end{figure}
\begin{figure}[!t]
\centering
\includegraphics [width=1.0\columnwidth] {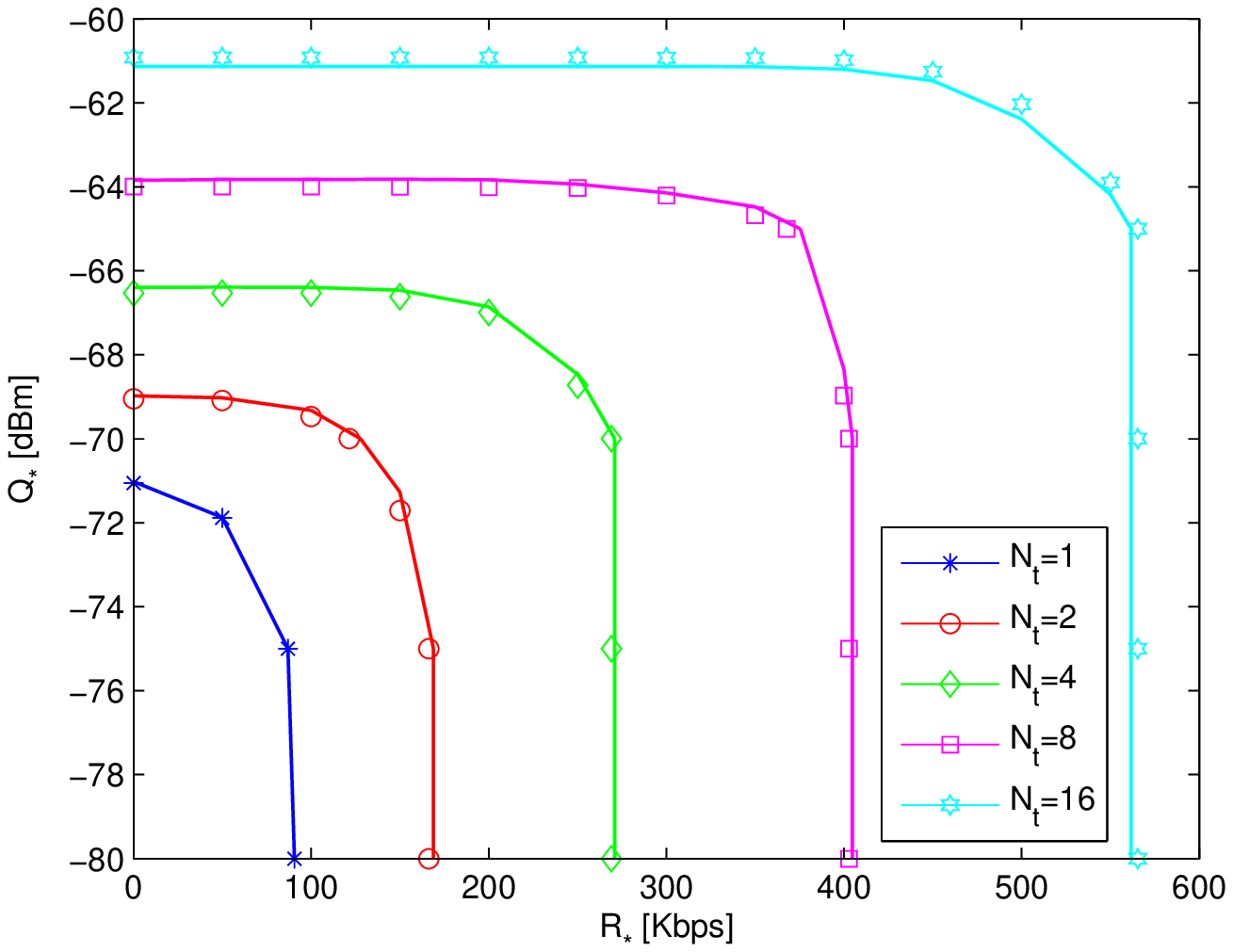}
\vspace{-0.75cm} \caption{Contour lines of the J-CCDF as a function of $N_t$ that correspond to the pairs $\left( {{\mathcal{R}}_* ,{\mathcal{Q}}_* } \right)$ such that $F_c \left( {{\mathcal{R}}_* ,{\mathcal{Q}}_* } \right) = 0.75$. The solid lines show the mathematical framework in \eqref{Eq_16} and the markers show Monte Carlo simulations. Setup: $N_r=2$ and $\rho=0.5$.} \label{Fig_2} \vspace{-0.25cm}
\end{figure}
\begin{figure}[!t]
\centering
\includegraphics [width=1.0\columnwidth] {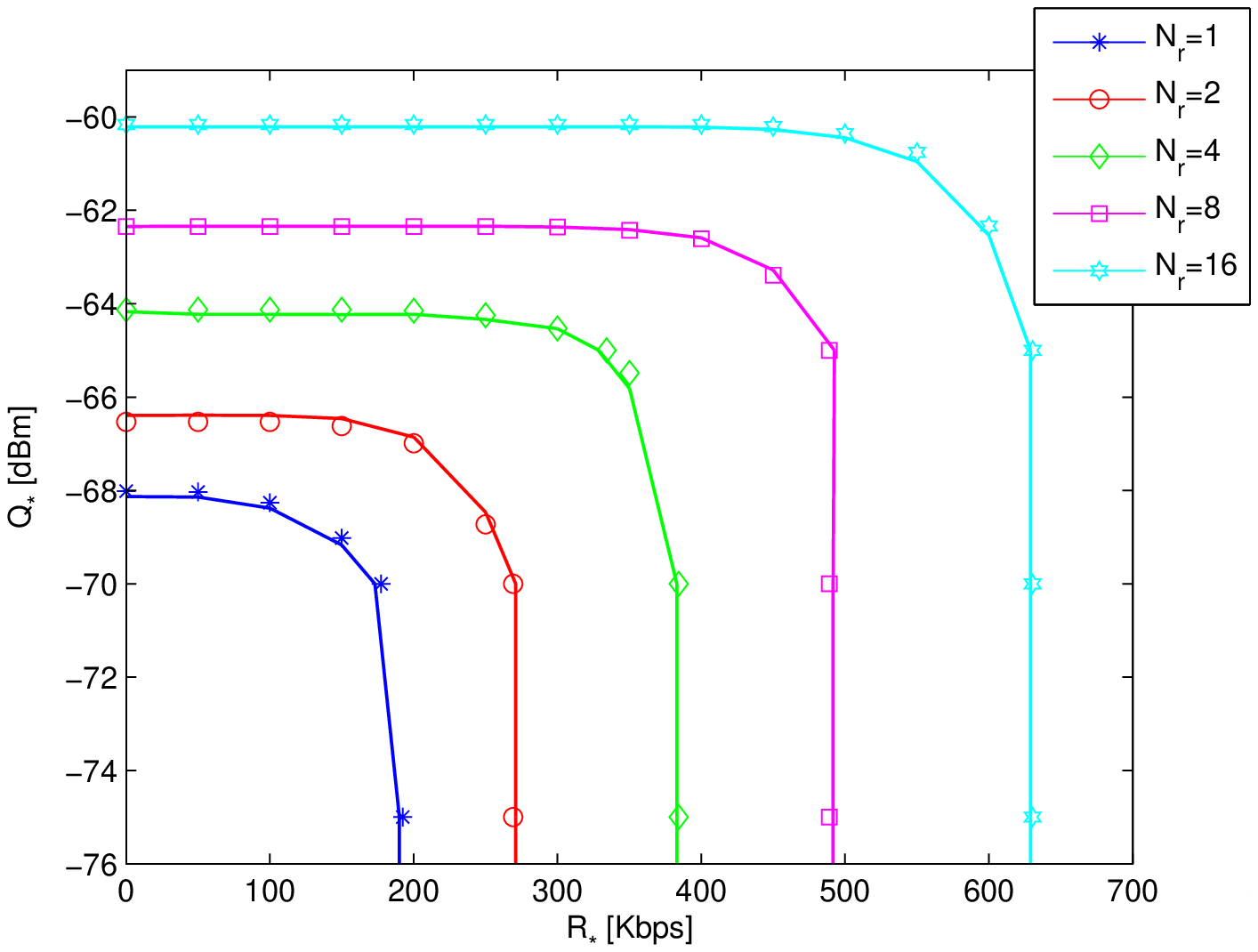}
\vspace{-0.75cm} \caption{Contour lines of the J-CCDF as a function of $N_r$ that correspond to the pairs $\left( {{\mathcal{R}}_* ,{\mathcal{Q}}_* } \right)$ such that $F_c \left( {{\mathcal{R}}_* ,{\mathcal{Q}}_* } \right) = 0.75$. The solid lines show the mathematical framework in \eqref{Eq_16} and the markers show Monte Carlo simulations. Setup: $N_t=4$ and $\rho=0.5$.} \label{Fig_3} \vspace{-0.25cm}
\end{figure}
\begin{figure*}[!t] 
\setcounter{equation}{17}
\begin{equation}
\label{Eq_18}
\begin{split}
 F_c \left( {{\mathcal{R}}_* ,{\mathcal{Q}}_* } \right)  & = \Pr \left\{ {{\mathcal{R}} \ge {\mathcal{R}}_* ,{\mathcal{Q}} \ge {\mathcal{Q}}_* } \right\} = \Pr \left\{ {{\mathcal{I}} \le {{\chi ^{\left( 0 \right)} r_* } \mathord{\left/
 {\vphantom {{\chi ^{\left( 0 \right)} r_* } {L^{\left( 0 \right)} }}} \right.
 \kern-\nulldelimiterspace} {L^{\left( 0 \right)} }} - {{\sigma _*^2 } \mathord{\left/
 {\vphantom {{\sigma _*^2 } P}} \right.
 \kern-\nulldelimiterspace} P},{\mathcal{I}} \ge  - {{\chi ^{\left( 0 \right)} } \mathord{\left/
 {\vphantom {{\chi ^{\left( 0 \right)} } {L^{\left( 0 \right)} }}} \right.
 \kern-\nulldelimiterspace} {L^{\left( 0 \right)} }} + {{q_* } \mathord{\left/
 {\vphantom {{q_* } P}} \right.
 \kern-\nulldelimiterspace} P}} \right\} \\
 & = \left\{ \begin{array}{l}
 \Pr \left\{ { - {{\chi ^{\left( 0 \right)} } \mathord{\left/
 {\vphantom {{\chi ^{\left( 0 \right)} } {L^{\left( 0 \right)} }}} \right.
 \kern-\nulldelimiterspace} {L^{\left( 0 \right)} }} + {{q_* } \mathord{\left/
 {\vphantom {{q_* } P}} \right.
 \kern-\nulldelimiterspace} P} \le {\mathcal{I}} \le {{\chi ^{\left( 0 \right)} r_* } \mathord{\left/
 {\vphantom {{\chi ^{\left( 0 \right)} r_* } {L^{\left( 0 \right)} }}} \right.
 \kern-\nulldelimiterspace} {L^{\left( 0 \right)} }} - {{\sigma _*^2 } \mathord{\left/
 {\vphantom {{\sigma _*^2 } P}} \right.
 \kern-\nulldelimiterspace} P}} \right\}\quad {\rm{if}}\quad \chi ^{\left( 0 \right)}  \ge \left( {{{{\mathcal{T}}_* } \mathord{\left/
 {\vphantom {{{\rm{T}}_* } P}} \right.
 \kern-\nulldelimiterspace} P}} \right)L^{\left( 0 \right)}  \\
 0\quad \hspace{7.77cm} {\rm{otherwise}} \\
 \end{array} \right. \\
 & = {\mathbb{E}}_{L^{\left( 0 \right)} } \left\{ {\int\nolimits_{\left( {{{{\mathcal{T}}_* } \mathord{\left/
 {\vphantom {{{\rm{T}}_* } P}} \right.
 \kern-\nulldelimiterspace} P}} \right)L^{\left( 0 \right)} }^{ + \infty } {F_{\mathcal{I}} \left( {\left. {x\left( {{{r_* } \mathord{\left/
 {\vphantom {{r_* } {L^{\left( 0 \right)} }}} \right.
 \kern-\nulldelimiterspace} {L^{\left( 0 \right)} }}} \right) - {{\sigma _*^2 } \mathord{\left/
 {\vphantom {{\sigma _*^2 } P}} \right.
 \kern-\nulldelimiterspace} P}} \right|L^{\left( 0 \right)} } \right)f_{\chi ^{\left( 0 \right)} } \left( x \right)dx} } \right\} \\ & - {\mathbb{E}}_{L^{\left( 0 \right)} } \left\{ {\int\nolimits_{\left( {{{{\mathcal{T}}_* } \mathord{\left/
 {\vphantom {{{\mathcal{T}}_* } P}} \right.
 \kern-\nulldelimiterspace} P}} \right)L^{\left( 0 \right)} }^{ + \infty } {F_{\mathcal{I}} \left( {\left. { - {x \mathord{\left/
 {\vphantom {x {L^{\left( 0 \right)} }}} \right.
 \kern-\nulldelimiterspace} {L^{\left( 0 \right)} }} + {{q_* } \mathord{\left/
 {\vphantom {{q_* } P}} \right.
 \kern-\nulldelimiterspace} P}} \right|L^{\left( 0 \right)} } \right)f_{\chi ^{\left( 0 \right)} } \left( x \right)dx} } \right\}
 \end{split}
\end{equation}
\normalsize \hrulefill \vspace*{-0pt}
\end{figure*}
\section{Numerical and Simulation Results} \label{Results}
In this section, we validate the mathematical framework in \eqref{Eq_16} against Monte Carlo simulations and study the impact of having multiple antennas at the BSs and at the LeDs. Further information on how Monte Carlo simulation results are obtained can be found in \cite{ACM_MDR}. Unless otherwise stated, the following setup is considered: $\nu  = {{c_0 } \mathord{\left/ {\vphantom {{c_0 } {f_c }}} \right. \kern-\nulldelimiterspace} {f_c }}$, where $c_0$ is the speed of light in m/sec and $f_c  = 2.1$ GHz is the carrier frequency; $\sigma _{{\rm{ID}}}^2  =  - 70$ dBm; $\sigma _N^2  =  - 174 + 10\log _{10} \left( {{{B}}_{{w}} } \right) + {\mathcal{F}}_N$ dBm, where ${{{B}}_{{w}} } = 200$ KHz and ${\mathcal{F}}_N = 10$ dB is the noise figure; $P=30$ dBm; $\zeta = 0.8$. The channel model and $\lambda$ are in agreement with \cite{ACM_MDR}: $D = 109.8517$ m, $q_{\text{LOS}}^{\left[ 0,D \right]} = 0.7195$, $q_{\text{LOS}}^{\left[ D,\infty \right]} = 0.0002$, $\beta_{\text{LOS}} = 2.5$, $\beta_{\text{NLOS}} = 3.5$, $\lambda = 1/(\pi R_{\text{cell}}^2)$ where $R_{\text{cell}} = 83.4122$ m is the average cell radius.

In Fig. \ref{Fig_1}, we illustrate the impact of $\rho$ on the achievable information rate and harvested power. The figure clearly demonstrates that $\rho$ needs to be optimized as a function of the pair $\left( {{\mathcal{R}}_* ,{\mathcal{Q}}_* } \right)$ being considered. This system-level optimization allows the LeDs to achieve a higher information rate and to harvest more power. The gap with respect to the ideal setup where information and power can be decoded and harvested simultaneously, respectively, is, however, non-negligible. This implies that practical receiver schemes, different from PS, are needed for further improving the performance of SWIPT-enabled cellular networks.

In Figs. \ref{Fig_2} and \ref{Fig_3}, we study the impact of the number of antennas at the BSs and at the LeDs, respectively. Both figures clearly demonstrate that multiple-antenna transmission is a promising technique, not only for enhancing the information rate but for increasing the harvested power as well.
\section{Conclusion} \label{Conclusion}
In this paper, we have proposed a mathematical framework for the analysis of SWIPT-enabled MIMO cellular networks. The accuracy of the proposed approach has been validated against Monte Carlo simulations. The numerical results have demonstrated that the use of multiple antennas at both the BSs and at the LeDs is capable of enhancing the achievable rate and increasing the harvested power simultaneously.
\section*{Acknowledgment}
This work is supported by the EC through the H2020-ETN-5Gwireless project (grant 641985) and the EPSRC through the Spatially Embedded Networks project (grant EP/N002350/1).
\appendices
\section{Proof of Proposition \ref{JCCDF__Prop}} \label{Appendix}
From \eqref{Eq_8} and \eqref{Eq_9}, the equalities in \eqref{Eq_18} shown at the top of this page hold. The proof follows by using the Gil-Pelaez inversion theorem \cite{GilPelaez} as follows:
\setcounter{equation}{18}
\begin{equation}
\label{Eq_19}
\begin{split}
& F_{\mathcal{I}} \left( {\left. z \right|L^{\left( 0 \right)} } \right) = 1/2 \\ & - \int\nolimits_0^{ + \infty } {\left( {\pi \omega } \right)^{ - 1} {\mathop{\rm Im}\nolimits} \left\{ {\exp \left( { - j\omega z} \right)\Phi _{\mathcal{I}} \left( {\left. \omega  \right|L^{\left( 0 \right)} } \right)} \right\}d\omega }
 \end{split}
\end{equation}
\noindent and the law of the unconscious statistician, \textit{i.e.}, ${\mathbb{E}}_{L^{\left( 0 \right)} } \left\{ {g\left( {L^{\left( 0 \right)} } \right)} \right\} = \int\nolimits_0^{ + \infty } {g\left( y \right)f_{L^{\left( 0 \right)} } \left( y \right)dy}$.

More precisely, the integral with respect to $x$ is computed by inserting \eqref{Eq_5} in \eqref{Eq_18} and by using the notable integral:
\begin{equation}
\label{Eq_20}
\int\nolimits_{\mathcal{A}}^{ + \infty } {x^u \exp \left( { - \mathcal{Z}x} \right)dx}  = \mathcal{Z}^{ - \left( {1 + u} \right)} \Gamma \left( {1 + u,\mathcal{AZ}} \right)
\end{equation}
\end{document}